\documentclass[12pt,preprint]{aastex}

\shortauthors{Harko, Cheng \& Tang}
\begin{document}
\title{Nucleation of quark matter in neutron stars cores}

\author{T.~Harko$^1$, K.~S.~Cheng$^2$ and P.~S. Tang$^3$}
\affil{Department of Physics, The University of Hong Kong, Pokfulam Road, Hong Kong, P. R. China\\
\email{$^1$harko@hkucc.hku.hk, $^2$hrspksc@hkucc.hku.hk},
$^3$anisia@bohr.physics.hku.hk}

\begin{abstract}
We consider the general conditions of quark droplets formation in
high density neutron matter. The growth of the quark bubble
(assumed to contain a sufficiently large number of particles) can
be described by means of a Fokker-Planck equation. The dynamics of
the nucleation essentially depends on the physical properties of
the medium it takes place. The conditions for quark bubble
formation are analyzed within the frameworks of both dissipative
and non-dissipative (with zero bulk and shear viscosity
coefficients) approaches. The conversion time of the neutron star
to a quark star is obtained as a function of the equation of state
of the neutron matter and of the microscopic parameters of the
quark nuclei. As an application of the obtained formalism we
analyze the first order phase transition from neutron matter to
quark matter in rapidly rotating neutron stars cores, triggered by
the gravitational energy released during the spinning down of the
neutron star. The endothermic conversion process, via
gravitational energy absorption, could take place, in a very short
time interval, of the order of few tens seconds, in a class of
dense compact objects, with very high magnetic fields, called
magnetars.
\end{abstract}


\section{Introduction}

Since \citet{Wi84}, following early proposals by \citet{It70} and \citet{Bo71},
suggested that strange quark matter, consisting of $%
u$-, $d$- and $s$-quarks is energetically the most favorable state
of the matter, the problem of existence of strange quark stars has
been intensively investigated in the physical and astrophysical
literature. The possibility that some compact objects could be
strange stars remains an interesting and intriguing, but still
open question. \citet{Wi84} also proposed two ways of formation of
strange matter: the quark-hadron phase transition in the early
universe and conversion of neutron stars into strange ones at
ultrahigh densities. In the theories of strong interaction quark
bag models suppose that breaking of physical vacuum takes place
inside hadrons. As a result vacuum energy densities inside and
outside a hadron become essentially different, and the vacuum
pressure on the bag wall equilibrates the pressure of quarks, thus
stabilizing the system. If the hypothesis of the quark matter is
true, then some of neutrons stars could actually be strange stars,
built entirely of strange matter \citep{AlFaOl86,HaZdSch86}).
However, there are general arguments against the existence of
strange stars \citep{CaFr91}. For an extensive review of strange
star properties see \citet{ChDaLu98}.

Several mechanisms have been proposed for the formation of quark
stars. Quark stars are expected to form during the collapse of the
core of a massive star after the supernova explosion, as a result
of a first or second order phase transition, resulting in
deconfined quark matter \citep{Da}. The proto-neutron star core, or
the neutron star core, is a favorable environment for the
conversion of ordinary matter to strange quark matter \citep{ChDa}.
Another possibility is that some neutron stars in low-mass X-ray
binaries can accrete sufficient mass to undergo a phase transition
to become strange stars \citep{Ch96}. This mechanism has also been
proposed as a source of radiation emission for cosmological
$\gamma $-ray bursts \citep{Ch98a}. Some basic properties of
strange stars like mass, radius, cooling, collapse and surface
radiation have been also studied \citep{ChHa,ChHa03}.

The physical mechanisms of the transition from neutron matter to quark
matter in an astrophysical background have been studied within several
models. The first is due to \citet{Ol87}, who used a non-relativistic
diffusion model. As such, this is a slow combustion model, with the burning
front propagating at a speed of approximately $10$ m/$\sec $. This is
determined primarily by the rate at which one of the down quarks inside the
neutrons is converted, through a weak decay, to a strange quark: $%
d+u\rightarrow s+u$. The second method of describing the
conversion process was first suggested by \citet{HoBe88}, and analyzed in detail by \citet{LuBeVu94} and \citet{LuBe95},
who modelled the conversion as a detonation. In this case the
conversion rate is several orders of magnitude faster than that
predicted by the slow combustion model. This model is based on the
relativistic shock waves and combustion theory. But regardless of
the way in which the transformation occurs, an initial seed of
quark matter is needed to start the process.

Real neutron stars have two conserved charges- electric and
baryonic. Therefore, a neutron star has more than one independent
component and in this sense it is a ``complex" system
\citep{Gl02}. The characteristics of a first-order phase
transition, like the deconfinement transition, are very different
in the two cases. In the ``complex" system the conserved charges
can be shared by the two phases in equilibrium in different
concentrations in each phase. The mixed phase, formed from hadrons
and quarks cannot exist in a simple body in the presence of the
gravity because the pressure in that phase is a constant
\citep{Gl00}. This causes a discontinuity in the density
distribution in the star occurring at the radius where Gibbs
criteria are satisfied. The isospin symmetry energy in neutron
rich matter will exploit the degree of freedom of readjusting the
charges between hadronic and quark phases in equilibrium so as to
reduce the symmetry energy to the extent consistent with charge
conservation. Regions of hadronic matter will have a net positive
charge neutralized by a net negative charge on the quark matter
regions \citep{Gl00}. Coulomb repulsion will prevent the regions
of like charge to grow too large, and the surface energy will act
in the opposite sense in preventing them to become too small.
Consequently, the mixed phase will form a Coulomb lattice so as to
minimize the sum of Coulomb and surface interface energy at each
proportion of phases. As the quark phase becomes more abundant,
the droplets merge to form rods, rods merge to form slabs etc.
\citep{Gl00}. Hence the actual geometric phase of quark nuggets in
a neutron star evolves as a function of the pressure and of the
surface energy between dense nuclear matter and quark matter.

In order to describe the process of formation and evolution of
quark nuclei in the neutron stars one must use nucleation theory.
The goal of nucleation theory is to compute the probability that a
bubble or droplet of the $A$ phase appears in a system initially
in the $B$ phase near the critical temperature \citep{LaLi80}.
Homogeneous nucleation theory applies when the system is pure.
Nucleation theory is applicable for first-order phase transitions
when the matter is not dramatically supercooled or superheated. If
substantial supercooling or superheating is present, or if the
phase transition is second order, then the relevant dynamics is
spinodal decomposition \citep{ShMo01}.

The nucleation of strange quark matter inside hot, dense nuclear matter was
investigated, with the use of Zel'dovich's kinetic theory of nucleation, by
\citet{HoBeVu92}. By assuming that the newly
formed strange quark matter bubble can be described by a simple bag model
containing $N_{q}$ ultrarelativistic quarks, and by assuming that the time
evolution of the radius of the bubble is described by an equation of the
form $dr/dt=(r-R_{c})/\tau _{w}$, where $R_{c}$ is the critical radius of
the bubble and $\tau _{w}$ the weak interaction time scale, the nucleation
rate $\xi $ is given by
\begin{equation}
\xi =2.2\times 10^{-2}\left( T/\sigma \right) ^{1/2}N_{qc}^{3/4}\tau
_{w}^{-1}\exp \left( -3.1N_{qc}^{1/2}\sigma /T\right) ,
\end{equation}
where $N_{qc}\leq 300$ is the critical quark number, $\sigma $ is the
surface tension of the bubble and $n_{n}$ and $n_{q}$ are the particle
number densities in the neutron and quark phase, respectively \citep{HoBeVu92}%
. In this way a lower bound for the temperature of the nucleation, $T\geq
2.1MeV$, can be obtained. The effects of the curvature energy term on
thermal strange quark matter nucleation have been considered by \citet{Ho94}, who derived, within the same approach, the following expression for
the nucleation rate:
\begin{equation}
\xi =\frac{r_{c}^{2}n_{q}n_{n}}{4\pi \tau _{w}}\left( T/\sigma \right)
^{1/2}\exp \left( -4\pi R_{c}^{2}/3T\right) .
\end{equation}

All the effects of the curvature enter only through the value of
$R_{c}$. As a general conclusion one obtains the result that even
the curvature term acts against strange quark matter nucleation,
the physical temperature of a just born proto-neutron star is, in
any model, more than enough to drive an efficient boiling of the
neutron material (the observations of the neutrino flux from
SN1987A are consistent with an effective temperature of $T\sim 4$
MeV). The possibility of stable strange quark matter (both bulk
and quasi-bulk) at finite temperature, and some of its properties
have been investigated, within the framework of the dynamical
density-dependent quark mass model of confinement, by
\citet{Ch93}. The behavior of the surface tension and the
stability of quark droplets at $T\neq 0$ has also been discussed,
with reference to strangelet formation in ultra-relativistic
heavy-ion collisions.

A different approach to the problem of thermal nucleation was taken by
\citet{OlMa94}. They used the assumption (standard in the
theory of bubble nucleation in first-order phase transitions) that bubbles
form at a rate given by $R\approx T^{4}\exp \left( -W_{c}/T\right) $, with $%
W_{c}$ the minimum work required to form a bubble with radius $r_{c}$. As a
main result it follows that if the bag constant lies in the interval where
three-flavor but not two-flavor quark matter is stable at zero pressure and
temperature ($145$ MeV$\leq B^{1/4}\leq 163$ MeV), then all or parts of a
neutron star will be converted into strange matter, during the first second
of its existence. For bag constant above the stability interval, a partial
transformation is still possible \citep{OlMa94}.

\citet{He95} calculated the rate of formation of quark matter
droplets in neutron stars from a combination of bubble formation rates in
cold degenerate and high temperature matter, taking into account nuclear
matter calculations of the viscosity and thermal conductivity. The droplet
formation rate is that of \citet{LaTu73}, given by $%
I=\left( k/2\pi \right) \Omega _{0}\exp \left( -W_{c}/T\right) $, with $%
\Omega _{0}=\left( 2/3\sqrt{3}\right) \left( \sigma /T\right) ^{3/2}\left(
r_{c}/\xi _{q}\right) ^{4}$ the statistical prefactor, which measures the
phase-space volume of the saddle point around $R_{c}$ that the droplet has
to pass, in its way to the lower energy state. $\xi _{q}$ is the quark
correlation length. The dynamical prefactor determines the droplet growth
rate and is given by $k=\left( 2\sigma /\Delta w^{2}r_{c}^{3}\right) \left(
\lambda T+2\left( 4\eta /3+\zeta \right) \right) $, where $\Delta w$ is the
enthalpy difference and $\lambda ,\eta $ and $\zeta $ are the thermal
conductivity and the shear and bulk viscosities, respectively. The droplet
formation rate can be expressed as \citep{He95}
\begin{equation}\label{eqn}
I=\frac{\sigma _{20}^{7/2}\mu _{400}^{2}\eta _{50}}{\Delta P_{10}\Delta
w_{10}^{2}T_{10}^{3/2}}\exp \left( 185-134\frac{\sigma _{20}^{3}}{\Delta
P_{10}^{2}T_{10}}\right) \textrm{ s}^{-1}\textrm{km}^{-3},
\end{equation}
with $\mu _{400}$ is the quark chemical potential in units of
$400$ MeV.

The total number $N$ of droplets formed by nucleation is the
integrated rate over the volume of the neutron star and the time
after the nucleation process starts at the moment $t_0$,
$N=\int_{0}^{R}4\pi r^{2}dr\int_{t_{0}}^{\infty }I\left[ \Delta
p\left( r\right) ,T\left( t\right) \right]dt$, where $R$ is the
radius of the neutron star. The pressure and temperature depend
sensitively on the equation of state of the nuclear and quark
matter. To convert the core of the neutron star into quark matter
at least one droplet must be formed, i.e., $N>1$. Then, with the
use of Eq. (\ref{eqn}), the condition for the formation of at
least one droplet requires $\sigma \leq 24$ MeV fm$^{-2}\Delta
P_{c,10}^{2/3}T_{c,10}^{1/3}$, where $P_{c}$ and $T_{c}$ are the
core values of the pressure and temperature, respectively
\citep{He95}. The pressure and enthalpy difference depends
strongly on the equation of state of nuclear and quark matter and
hence the droplet formation rate cannot be reliably estimated.

The effect of subcritical hadron bubbles on an inhomogeneous first
order quark-hadron phase transition was studied in
\citet{ShMoGuGl00}. The transition from nuclear matter (consisting
of neutrons, protons and electrons) to matter containing
strangeness was considered, within a mean field type model and
with the use of Langer nucleation theory, by \citet{No02}. An
estimate of the time the new phase to appear at various densities
and times in the cooling history of a protoneutron star was also
given.

On the other hand, as a result of an increase of the central density of the
star, due for example to accretion or spinning down, a metastable,
super-compressed neutron phase can appear, with a star consisting of the new
phase surrounded by normal neutron matter \citep{Gr98}. This situation is
similar to a gas undergoing a phase transition to its liquid state if
compressed to a volume $V_{c}$ at fixed $T$. If it is slowly compressed, it
may stay in the vapor state even for $V<V_{c}$ (similar examples are liquid
water freed of dissolved air which may be heated above boiling temperature
without the formation of vapor and cooled below freezing temperature without
solidification). The transition from the super-compressed neutron phase to
the quark phase can take place in a very long time, which in some cases can
be longer than the estimated life of the neutron star \citep{Gr98}. Therefore an increase in the density of the neutron star does not
automatically lead to the transition to quark matter inside the star.

The effect of the magnetic field on the quark star structure and
on the nucleation process of the quark bubbles has been studied in
\citep{Ch95,ChSa96,Ch96a}. In the presence of strong magnetic
fields the equation of state of strange quark matter changes
significantly. The strange stars become more compact, the magnetic
field reducing the mass and radius of the star. The surface energy
and the curvature term of the quark phase both diverge, with the
surface tension diverging logarithmically, while the curvature
term diverges much faster. Therefore the thermal nucleation of
quark bubbles in a compact metastable state of neutron matter is
completely forbidden in the presence of a strong magnetic field.

These results for the formation of quark bubbles in neutron matter
have been obtained by using the Csernai and Kapusta theory of
nucleation \citep{CsKa92,CsKa92b} and its extension including the
thermal conductivity of dense matter \citep{VeVi94}. In this
theory both the hadron and quark materials are considered as
substances with zero baryonic number, and are treated by using a
relativistic formalism. The principal result in this approach is
the suggestion that the prefactor $k$ is proportional to the
transport coefficients (viscosity and thermal conductivity) of the
neutron matter, and thus when these coefficients vanish a quark
bubble necessarily does not form.

However, \citet{RuFr96} argued that the energy flow does
not vanish in absence of any heat conduction or viscous damping. Since the
change of energy density $e$ in time is given, in the low velocity limit, by
the conservation equation $\partial e/\partial t=-\nabla \left( w\vec{v}%
\right) $ \citep{CsKa92}, where $w$ is the enthalpy and $\vec{v}$ the
velocity, this implies that the energy flow $\sim w\vec{v}$ is always
present. Therefore, an expression for the prefactor can be derived, which
does not vanish in the absence of viscosity. The viscous effects cause only
a small perturbation to the prefactor. The differences between the
Csernai-Kapusta (CK) \citep{CsKa92} and Ruggieri-Friedman (RF) \citep{RuFr96}
results are due to the technical differences in the treatment of the
pressure gradients.

A generalized approach, following the CK formalism leading to the prefactor
in both viscous and non-viscous regimes was developed by \citet{ShMoGu01}.
Unlike in \citet{CsKa92}, the linearized relativistic
hydrodynamics equations have been solved in all regions.

It is the purpose of the present paper to extend the Zel'dovich
nucleation theory for the formation of quark droplets in neutron
matter, by taking into account both the effects of the energy flow
and of thermal conductivity, shear and bulk viscosities. As a
result a more general expression for the quark matter droplet rate
formation can be obtained, which is also valid in the limiting
case of vanishing conductivity and viscosity coefficients. In the
nucleation theory of \citet{CsKa92}, the transition from the
neutron to quark state is possible only for a normal matter having
viscous properties. For zero bulk and shear viscosities the
transition is impossible. This assumption is, however, to
restrictive, and the transition can also take place for a perfect
(no viscosity or heat conduction) neutron matter fluid. We also
derive the expression of the rate formation for this case.

An alternative point of view of the transformation of neutron
matter into strange matter can be developed if one assumes that
the conversion process is endothermic. In this case the strange
quark matter formation inside a neutron star is triggered by an
external source of energy (accretion from the companion star or
spin-down). Pulsars are born with an enormous angular momentum and
rotational energy, which they radiate over a long period of time
via electromagnetic radiation and electron-positron pair emission.
When rotating rapidly, a pulsar is centrifugally flattened. With
decreasing angular velocity the central density of the star is
increasing and may attain the critical density necessary for a
phase transition. First at the center and then in an expanding
region the neutron matter will be converted to a highly
compressible quark matter phase \citep{GlPeWe97}. The conversion
of neutron matter to quark matter alters the moment of inertia of
the star, and the epoch over which conversion takes place will be
signaled in the spin-down characteristics of the pulsar. A
measurable quantity, the braking index $\Omega
\ddot{\Omega}/\dot{\Omega}^{2}$, where $\Omega $ is the angular
velocity of the star, could be an observational indicator of the
slow transition from neutron to quark phase \citep{GlPeWe97}. By
using the formalism developed for the nucleation of quark matter
we reconsider the idea of phase transition from neutron matter to
quark matter in rotating compact stellar objects, by assuming that
the nucleation and the growth of the quark droplets and the
formation of a quark core in a neutron star is a result of the
transfer of the gravitational energy to the quark droplets, due to
the spinning down of the rapidly rotating neutron star.

The present paper is organized as follows. The nucleation kinetics
of strange matter bubbles is considered, within the framework of
the Zel'dovich nucleation theory, in Section II. The expressions
for the rate of formation of quark matter are derived, in both the
perfect and dissipative neutron fluid case, in Section III. The
transition of a neutron star to a quark star, due to increase of
the central density as a result of the spinning down is considered
in Section IV. In Section V we discuss and conclude our results.

\section{Nucleation kinetics of strange matter in neutron stars}

We consider that the change from the metastable neutron phase to the stable
quark phase occurs as the result of fluctuations in a homogeneous medium,
formed of neutrons, in which small quantities of the quark phase (called
bubbles or nuclei) are randomly generated. Since the process of creation of
an interface is energetically unfavorable, it follows that when a quark
nucleus is below a certain size, it is unstable and disappears again.
Surface effects disfavor the survival of small bubbles below the radius $%
R_{c}$ (called critical size-the nuclei of this size are called critical
nuclei or bubbles), which is nothing but the value that extremizes the
thermodynamical work $W$ necessary to create the bubbles. Only nuclei whose
size $r$ is above the value $R_{c}$ are stable, and can survive \citep{LaLi80}%
.

The nuclei are assumed to be macroscopic objects containing a large number
of particles (quarks).

For strongly degenerate neutron matter the thermodynamic work $W$ necessary
to create a quark bubble is \citep{AlOl89,MaSv91,MaOl91,OlMa94,Ho94}
\begin{equation}
W=\left[ n_{q}\left( \mu _{q}-\mu _{n}\right) -\left( P_{q}-P_{n}\right) %
\right] \frac{4\pi }{3}r^{3}+4\pi \sigma r^{2}+8\pi \gamma r+E_{c},
\end{equation}
where $P_{q,n}$ are the pressures of the quark matter and neutron matter
(exterior to the bubble, assumed to be spherical), $\sigma $ is the surface
tension, $n_{q}$ is the particle-number density of quark matter, $\mu _{q,n}$
are the chemical potentials of each phase, $\gamma $ is the curvature
coefficient and $E_{c}=\left( 3/5\right) Z^{2}e^{2}/r$ is the Coulomb energy
of the droplet \citep{HePeSt93a}. Here $Ze=\left( \rho _{q}-\rho _{n}\right)
V_{d}$ is the excess charge of the droplet with volume $V_{d}$ compared with
the surrounding medium, and $\rho _{q}$ and $\rho _{n}$ are the charge
densities of the quark and nuclear matter, respectively. Therefore the
Coulomb energy of the droplet is given by $E_{c}=16\pi ^{2}\left( \rho
_{q}-\rho _{n}\right) ^{2}r^{5}/15$. Assuming that the quark matter is
immersed in a uniform background of electros, and that it is in $\beta $
equilibrium, with $\mu _{d}=\mu _{s}=\mu _{u}+\mu _{e}$, it follows that the
total electric charge in the quark phase can be expressed as $\rho
_{q}\approx e\mu _{q}\left( m_{s}^{2}/2-2\mu _{e}\mu _{q}\right) /\pi ^{2}$,
where $\mu _{u}\approx \mu _{d}\equiv \mu _{q}$ \citep{HePeSt93a}. Generally $%
\rho _{q}>>\rho _{n}$.

The appropriate form of the curvature energy can be obtained from \citep{Ma93,HePeSt93a}
\begin{equation}
E_{curv}=\frac{gr}{3\pi }\int_{0}^{\infty }dkk\left[ 1+\exp \left( \left(
k-\mu /T\right) \right) \right] ^{-1}=\frac{g\mu ^{2}r}{6\pi }\left[
1+O\left( T/\mu \right) ^{2}\right] ,
\end{equation}
where $g$ is the statistical weight and $\mu $ the chemical potential. Hence
the curvature coefficient is given by $\gamma =3\mu ^{2}/8\pi ^{2}\sim 18$
MeVfm$^{-1}\left( \mu /300\textrm{ MeV}\right) ^{2}$ \citep{HePeSt93a}.

Requiring $W$ to be an extreme, $\partial W/\partial r=0$, gives the
following equation for the value of the critical radius $r=R_{c}$:
\begin{equation}
4\pi \left[ n_{q}\left( \mu _{q}-\mu _{n}\right) -\left( P_{q}-P_{n}\right) %
\right] r^{2}+8\pi \sigma r+8\pi \gamma =-\frac{16\pi ^{2}}{3}\left( \rho
_{q}-\rho _{n}\right) ^{2}r^{4}.  \label{eqgen}
\end{equation}

By neglecting the Coulomb energy, we obtain for the critical radius $R_{c0}$
of the bubble the expression
\begin{equation}
R_{c0}=\frac{\sigma }{C}\left( 1+\sqrt{1+b_{0}}\right) ,  \label{rc0}
\end{equation}
where
\begin{equation}
C=-n_{q}\left( \mu _{q}-\mu _{n}\right) +\left( P_{q}-P_{n}\right)
=-n_{q}\Delta \mu +\Delta P>0,
\end{equation}
and
\begin{equation}
b_{0}=2C\left| \gamma \right| /\sigma ^{2}>0.
\end{equation}

The parameter $b_{0}$ can also be represented as $b_{0}\approx 6\left( C/20%
\textrm{ MeVfm}^{-3}\right) \left( \gamma /18\textrm{ MeVfm}^{-1}\right) $ $\left[
\sigma /\left( 75\textrm{ MeV}\right) ^{3}\right] ^{-2}$ \citep{Ho94}.

To find the value of the critical radius for a non-negligible electrostatic
energy of the bubble we use an iterative method, by substituting in the
right-hand side of Eq.~(\ref{eqgen}) $r$ by the zero'th order approximation
given by Eq.~(\ref{rc0}). Then, in the first order, the critical radius is
given by
\begin{equation}
R_{c}=\frac{\sigma }{C}\left( 1+\sqrt{1+b}\right) ,  \label{rc}
\end{equation}
where
\begin{equation}
b=\frac{C}{\sigma ^{2}}\left[ 2\gamma +\frac{4\pi }{3}\left( \rho _{q}-\rho
_{n}\right) ^{2}R_{c0}^{4}\right] .
\end{equation}

For values of $C=20$ MeVfm$^{-3}$, $\gamma =18$ MeVfm$^{-1}$, $\sigma $ in
the range $\left( 75-100\textrm{ MeV}\right) ^{3}$, $\rho _{q}\approx -0.4e$ fm%
$^{-3}$ and negligible $\rho _{n}$ the critical radius is of the order of $%
R_{c}=2-7$ fm. The Coulomb correction has the effect of increasing the
critical radius. The value $r=R_{c}$ corresponds to the limit beyond which
large quantities of the quark phase begin to be formed. In fact, it is more
appropriate to refer not to a limit point $r=R_{c}$, but to a critical range
of values of $r$ near that point, with width $\delta r\sim \left( T/4\pi
\sigma \right) ^{1/2}$ \citep{LaLi80}. The fluctuational development of
nuclei in this range can still, with high probability, throw them back into
the subcritical range, but nuclei beyond the critical range will inevitably
develop into the new quark phase.

The minimum critical work required to form a stable quark bubble is
therefore given by
\begin{equation}
W_{c}=\frac{4\pi \sigma ^{3}}{3C^{2}}F(b)\left[ 1+4\pi \left( \rho _{q}-\rho
_{n}\right) ^{2}\frac{\sigma ^{2}}{C^{3}}G(b,b_{0})\right] ,
\end{equation}
where we denoted
\begin{equation}
F(b)=2+2\left( 1+b\right) ^{3/2}+3b,
\end{equation}
and
\begin{equation}
G(b,b_{0})=\frac{\left( 1+\sqrt{1+b}\right) ^{5}}{2+2\left( 1+b\right)
^{3/2}+3b}\left[ \frac{1}{5}-\frac{\left( 1+\sqrt{1+b_{0}}\right) ^{4}}{%
\left( 1+\sqrt{1+b}\right) ^{4}}\right] .
\end{equation}

Within a purely thermodynamic approach, one can put only the problem of
calculating the probability of occurrence in a medium of fluctuational
nuclei of various sizes, the medium being regarded as in equilibrium.
Instead of the thermodynamic probability of nucleation, it is more
convenient to use the equilibrium distribution function for nuclei of
various sizes existing in the medium, denoted $f_{0}(r)$. $f_{0}dr$ is the
number of nuclei per unit volume of the medium with sizes in the range $dr$.
According to the thermodynamic theory of fluctuations \citep{LaLi80},
\begin{equation}
f_{0}\left( r\right) \sim \exp \left[ -W(r)/T\right] .
\end{equation}

Near $r=R_{c}$ the thermodynamic work can be expressed as
\begin{eqnarray}
W\left( r\right)&=&\frac{4\pi \sigma ^{3}}{3C^{2}}F(b)\left[ 1+\frac{24\pi }{5%
}\left( \rho _{q}-\rho _{n}\right) ^{2}\frac{\sigma ^{2}}{C^{3}}G(b)\right] \nonumber\\
&&-%
\left[ \frac{12\pi \sigma C^{3}\sqrt{1+b}-32\pi ^{2}\left( \rho _{q}-\rho
_{n}\right) ^{2}\sigma ^{3}\left( 1+\sqrt{1+b}\right) ^{3}}{3C^{3}}\right]
\left( r-R_{c}\right) ^{2}.
\end{eqnarray}

Therefore for the equilibrium distribution function we find the expression
\begin{equation}
f_{0}(r)=f_{0}\left( R_{c}\right) \exp \left[ \frac{12\pi \sigma C^{3}\sqrt{%
1+b}-32\pi ^{2}\left( \rho _{q}-\rho _{n}\right) ^{2}\sigma ^{3}\left( 1+%
\sqrt{1+b}\right) ^{3}}{3C^{3}T}\left( r-R_{c}\right) ^{2}\right] ,
\end{equation}
where $f_{0}\left( R_{c}\right) =C_{0}\times \exp \left( -\frac{W_{c}}{T}%
\right) $, $C_{0}=$constant.

In order to estimate the coefficient $C_{0}$ of the exponential in $%
f_{0}\left( R_{c}\right) $ we follow \citet{LaLi80}, and
assume $C_{0}=n_{q}n_{n}R_{c}^{2}$, with $n_{n}$ and $n_{q}$ the particle
number densities in the neutron and quark phases, respectively. Thus we
obtain
\begin{equation}
f_{0}\left( R_{c}\right) =n_{q}n_{n}R_{c}^{2}\exp \left\{ -\frac{4\pi \sigma
^{3}}{3C^{2}T}F(b)\left[ 1+4\pi \left( \rho _{q}-\rho _{n}\right) ^{2}\frac{%
\sigma ^{2}}{C^{3}}G(b,b_{0})\right] \right\} .
\end{equation}

Let $f(t,r)$ be the kinetic size distribution function of the nuclei. The
elementary process which changes the size of a nucleus is the attachment to
it, or the loss by it, of a quark droplet, and this is to be regarded as a
small change, since the nuclei are considered to be macroscopic objects.
Therefore the growth of the nuclei is described by a Fokker-Planck equation
\citep{LaLi80}
\begin{equation}
\frac{\partial f\left( t,r\right) }{\partial t}=-\frac{\partial j}{\partial r%
},
\end{equation}
where $j=-B\frac{\partial f}{\partial r}+Af$ is the flux in the size space. $%
B$ is the nuclear size diffusion coefficient and $A$ is connected with $B$
by a relationship, which follows from the fact that for an equilibrium
distribution $j=0$. Therefore we find $A=-BW^{\prime }(r)/T$.

In the case of a continuous stationary phase-transition process we have $j=$%
constant. The constant flux is just the number of nuclei passing through the
critical range per unit time per unit volume of the medium, i.e. it defines
the rate of the process. With the use of the condition of the constant flux
we obtain $-Bf_{0}\frac{\partial }{\partial r}\left( \frac{f}{f_{0}}\right)
=j$ giving
\begin{equation}
\frac{f}{f_{0}}=-j\int \frac{dr}{Bf_{0}}+const.
\end{equation}

The constant in this equation and $j$ are found from the boundary conditions
for small and large $r$. The fluctuation probability increases rapidly with
decreasing size and small nuclei have a high probability of occurrence. This
is expressed by the boundary condition $f/f_{0}\rightarrow 1$ as $%
r\rightarrow 0$. The boundary condition for large $r$ can be established by
noting that above the critical range the function $f_{0}$ increases without
limit, whereas the true distribution function $f(r)$ remains finite. This
situation is expressed by imposing the boundary condition $f/f_{0}=0$ for $%
r\rightarrow \infty $. The solution which satisfies the above conditions is
\citep{LaLi80}
\begin{equation}
\frac{f}{f_{0}}=j\int_{r}^{\infty }\frac{dr}{Bf_{0}},\frac{1}{j}%
=\int_{0}^{\infty }\frac{dr}{Bf_{0}}.
\end{equation}

In these equations the integrand has a sharp maximum at $r=R_{c}$. By
extending the integration with respect to $r$ from $-\infty $ to $+\infty $,
one obtain for the number of viable quark nuclei formed in stationary
conditions per unit time and per unit volume the expression:
\begin{equation}
j=\sqrt{\frac{12\sigma C^{3}\sqrt{1+b}-32\pi \left( \rho _{q}-\rho
_{n}\right) ^{2}\sigma ^{3}\left( 1+\sqrt{1+b}\right) ^{3}}{3C^{3}T}}B\left(
R_{c}\right) f_{0}\left( R_{c}\right) .
\end{equation}

Above the critical range, the distribution function is constant: having
reached that point, the nucleus becomes steadily larger, with practically no
change in the reverse direction. Accordingly we can neglect the term
containing the derivative $\partial f/\partial r$ in the flux, leading to $%
j=Af$. From the significance of the flux it follows that the coefficient $A$
acts as a velocity in size space, $A=\left( dr/dt\right) _{macro}$ \citep{LaLi80}. Therefore we find for $B$:
\begin{eqnarray}
B&=&-\frac{T}{W^{\prime }(r)}\left( \frac{dr}{dt}\right) _{macro}\nonumber\\
&=&\frac{3C^{3}T%
}{2\left[ 12\pi \sigma C^{3}\sqrt{1+b}-32\pi ^{2}\left( \rho _{q}-\rho
_{n}\right) ^{2}\sigma ^{3}\left( 1+\sqrt{1+b}\right) ^{3}\right] \left(
r-R_{c}\right) }\left( \frac{dr}{dt}\right) _{macro}.
\end{eqnarray}

The rate of growth of the bubble radius near the critical radius is given by
$\left( \frac{dr}{dt}\right) _{macro}=k\left( r-R_{c}\right) $, where $k$ is
the dynamical prefactor \citep{LaTu73,CsKa92}.

The prefactor $k$ has been evaluated, by solving the equations of
relativistic fluid dynamics in all regions, in \citet{ShMoGu01}. The result
(also taking into account heat conduction) is
\begin{equation}
k=\sqrt{\frac{2\sigma }{R_{c}^{3}}\frac{w_{n}}{\left( \Delta w\right) ^{2}}}+%
\frac{1}{c_{s}^{2}}\frac{\sigma }{R_{c}^{3}\left( \Delta w\right) ^{2}}\left[
\lambda _{n}T+2\left( \frac{4}{3}\eta _{n}+\xi _{n}\right) \right] ,
\label{pre}
\end{equation}
where $c_{s}$ is a constant (the velocity of the sound in the medium around
the saddle configuration) and $\lambda _{n}$, $\eta _{n}$ and $\xi _{n}$ are
the thermal conductivity and shear and bulk viscosity coefficients of the
neutron matter, respectively. The first term in the above equation is the
same as obtained by \citet{RuFr96}, corresponding to the
case of non-viscous matter. The second term is similar to the results
obtained by \citet{CsKa92} and \citet{VeVi94}, except with
a minor difference , i.e. instead of $4$ there is a factor $c_{s}^{2}$ in
the numerator.

\section{Neutron matter to quark matter transition rates}

With the use of Eq.~(\ref{pre}) for the prefactor it follows that the rate
of formation of quark bubbles in neutron matter is given by
\begin{eqnarray}
j &=&\frac{n_{q}n_{n}R_{c}^{2}}{2\pi }\sqrt{\frac{3C^{3}T}{12\sigma C^{3}%
\sqrt{1+b}-32\pi \left( \rho _{q}-\rho _{n}\right) ^{2}\sigma ^{3}\left( 1+%
\sqrt{1+b}\right) ^{3}}}\nonumber \\
&&\left[ \sqrt{\frac{2\sigma }{R_{c}^{3}}\frac{w_{n}}{%
\left( \Delta w\right) ^{2}}}+\frac{\sigma \left( \lambda _{n}T+2\left(
\frac{4}{3}\eta _{n}+\xi _{n}\right) \right) }{c_{s}^{2}R_{c}^{3}\left(
\Delta w\right) ^{2}}\right] \times  \nonumber \\
&&\exp \left\{ -\frac{\frac{4\pi \sigma ^{3}}{3C^{2}}F(b)\left[ 1+4\pi
\left( \rho _{q}-\rho _{n}\right) ^{2}\frac{\sigma ^{2}}{C^{3}}G(b,b_{0})%
\right] }{T}\right\} .  \label{j}
\end{eqnarray}

To obtain the number $\xi $ of the net strange quark matter bubbles, the
rate $j$ must be multiplied by the time interval available for prompt
nucleation, $\Delta t$ and by the volume $V_{0}$, where the nucleation can
take place in the dense core \citep{HoBeVu92}. We impose that at least one
quark bubble appears (which would suffice to convert the whole neutron
star). Therefore, with the use of Eq.~(\ref{j}), one obtains the following
general condition for a quark bubble formation in a neutron star core:
\begin{equation}
\xi =j\Delta t\Delta V_{0}\geq 1.  \label{cond}
\end{equation}

The quark bubble formation rate essentially depends on the enthalpy
difference in the two phases. In the density range of interest, all the
neutron matter EOS can be very well parameterized as polytropes, $%
P_{n}=Kn_{B}^{\gamma }$ \citep{LuBeVu94}. The corresponding energy density in
the neutron phase is $\varepsilon _{n}=n_{B}m_{n}+\frac{1}{\gamma -1}%
Kn_{B}^{\gamma }$ , leading to $w_{n}=\varepsilon _{n}+P_{n}=n_{B}m_{n}+%
\frac{\gamma }{\gamma -1}Kn_{B}^{\gamma }$. We also assume that the formed
quark bubble consist of $u$ and $d$ quarks in the ratio $1:2$; only later
weak interactions may change the composition to an energetically more
favorable state. The quarks chemical potentials are related by $\mu
_{d}=2^{1/3}\mu _{u}$, and, assuming chemical equilibrium across the phase
boundary, we also have $\mu _{n}=\mu _{u}+2\mu _{d}=\left( 1+2^{4/3}\right)
\mu _{u}$ \citep{OlMa94}. Then the pressure in the quark phase is (assuming a
simple bag model) $P_{q}=\frac{\mu _{u}^{4}+\mu _{d}^{4}}{4\pi ^{2}}-B$ and $%
\varepsilon _{q}=3P_{q}+B$ \citep{ChDaLu98}, giving $w_{q}=\frac{\mu
_{u}^{4}+\mu _{d}^{4}}{\pi ^{2}}-3B$. Hence, the enthalpy difference in the
two phases can be approximated by
\begin{equation}
\Delta w=\frac{\mu _{u}^{4}+\mu _{d}^{4}}{\pi ^{2}}-3B-n_{B}m_{n}-\frac{%
\gamma }{\gamma -1}Kn_{B}^{\gamma }.
\end{equation}

Assuming that the energy flow is provided by the viscous effects only, one
obtains for the prefactor the expression \citep{VeVi94}
\begin{equation}
k=\frac{2\sigma }{\left( \Delta w\right) ^{2}R_{c}^{3}}\left[ \lambda
_{n}T+2\left( \frac{4}{3}\eta _{n}+\xi _{n}\right) \right] .  \label{k1}
\end{equation}

In the limit of zero baryon number, $\lambda _{n}\rightarrow 0$ and we
obtain the result of \citet{CsKa92}. If the matter is
baryon-rich, but viscous damping is negligible, $\eta _{n},\xi
_{n}\rightarrow 0$, and we obtain the results of \citet{LaTu73} and \ \citet{Ka75}.

Therefore the condition of the formation of a quark bubble is given by
\begin{eqnarray}  \label{cond1}
&&\frac{1}{\pi }\frac{n_{q}n_{n}}{\left( \Delta w\right) ^{2}R_{c}}\sqrt{%
\frac{3C^{3}T\sigma }{12C^{3}\sqrt{1+b}-32\pi \left( \rho _{q}-\rho
_{n}\right) ^{2}\sigma ^{2}\left( 1+\sqrt{1+b}\right) ^{3}}}\left[ \lambda
_{n}T+2\left( \frac{4}{3}\eta _{n}+\xi _{n}\right) \right] \times  \nonumber
\\
&&\exp \left\{ -\frac{4\pi \sigma ^{3}}{3C^{2}T}F(b)\left[ 1+4\pi \left(
\rho _{q}-\rho _{n}\right) ^{2}\frac{\sigma ^{2}}{C^{3}}G(b,b_{0})\right]
\right\} \Delta t\Delta V_{0}\geq 1.
\end{eqnarray}

The transport properties of dense matter have been intensively investigated
in both high energy physics and astrophysical frameworks \citep{FlIt76,FlIt79,FlIt81,HaMo93}. The shear viscosity $\eta $ and
thermal conductivity $\lambda $ for neutron matter have been derived by
\citet{Da84} from the Uhlenbeck-Uehling equation (see also \citet{Sa89} and \citet{CuLiSp90}). They are given by $\eta =\left( 16\pi
^{2}\right) ^{-1}\left( p_{F}^{5}/m_{n}^{2}\tilde{\sigma}_{1}\right) T^{-2}$
and $\lambda =5\left( p_{F}^{3}/m_{n}^{2}\tilde{\sigma}_{2}\right) T^{-1}/96$%
, \ where $p_{F}$ is the Fermi momentum and $\tilde{\sigma}_{1},\tilde{\sigma%
}_{2}$ are some quantities related to particle-particle cross
sections, estimated in \citet{Da84}. The contribution of the bulk
viscosity and thermal conductivity are negligible, and the main
contribution is from the shear viscosity, which is typically of
the order of $\eta \sim 50$ MeV/fm$^{2}$ \citep{He95}. But in the
case of an ideal neutron gas, with zero viscosity, there will be
no bubble growth, and in this case the transition from neutron
matter to quark matter in astrophysical objects cannot take place.

Assuming that the energy flow does not vanish in the absence of any heat
conduction or viscous damping, and considering that the viscous effects are
small and can be neglected, the condition of the formation of a quark bubble
inside the dense core of a neutron star becomes:
\begin{eqnarray}  \label{cond2}
&&\frac{1}{\sqrt{2}\pi }\frac{n_{q}n_{n}}{\Delta w}\sqrt{\frac{%
3C^{3}R_{c}Tw_{n}}{12C^{3}\sqrt{1+b}-32\pi \left( \rho _{q}-\rho _{n}\right)
^{2}\sigma ^{2}\left( 1+\sqrt{1+b}\right) ^{3}}}\times  \nonumber \\
&&\exp \left\{ -\frac{4\pi \sigma ^{3}}{3C^{2}T}F(b)\left[ 1+4\pi \left(
\rho _{q}-\rho _{n}\right) ^{2}\frac{\sigma ^{2}}{C^{3}}G(b,b_{0})\right]
\right\} \Delta tV_{0}\geq 1.
\end{eqnarray}

For $r>R_{c}$ the time growth of the radius of the quark droplets can be
described by the equation \citep{CsKa92,ShMoGu01}
\begin{equation}
\frac{dr}{dt}\approx kR_{c}^{2}\frac{r-R_{c}}{r^{2}},
\end{equation}
with the general solution
\begin{equation}
t\approx \frac{1}{k}\left[ \frac{1}{2}\left( \frac{r}{R_{c}}\right) ^{2}+%
\frac{r}{R_{c}}+\ln \frac{r-R_{c}}{R_{c}}-4\right] ,
\end{equation}
where we have used the initial condition $r(0)=2R_{c}$. Therefore for a
neutron star with radius $R$, the time $t_{conv}$ required for the
conversion of the whole star to a quark star can be obtained from
\begin{equation}
t_{conv}\approx \frac{1}{k}\left[ \frac{1}{2}\left( \frac{R}{R_{c}}\right)
^{2}+\frac{R}{R_{c}}+\ln \frac{R-R_{c}}{R_{c}}-4\right] .
\end{equation}

Since $R>>R_{c}$, with a very good approximation we find
\begin{equation}
t_{conv}\approx \frac{1}{2k}\left( \frac{R}{R_{c}}\right) ^{2}.
\label{tconv}
\end{equation}

Neglecting the viscous effects, and by assuming again that the neutron
matter is described by a polytropic equation of state, while the quark phase
obeys the simple bag model equation of state, one obtains for the conversion
time
\begin{equation}
t_{conv}\approx \frac{\left( \Delta w\right) ^{2}}{\sqrt{8\sigma R_{c}w_{n}}}%
R^{2}=\frac{\left( \frac{\mu _{u}^{4}+\mu _{d}^{4}}{\pi ^{2}}-3B-n_{B}m_{n}-%
\frac{\gamma }{\gamma -1}Kn_{B}^{\gamma }\right) ^{2}}{\sqrt{8\sigma
R_{c}\left( n_{B}m_{n}+\frac{\gamma }{\gamma -1}Kn_{B}^{\gamma }\right) }}%
R^{2}.  \label{conv}
\end{equation}

The volume $V(t)$ of the neutron matter converted to quark matter after a
time interval $t<t_{conv}$ is given, in this simple model, by
\begin{equation}
V\left( t\right) \approx \frac{2^{17/4}\pi \left( \sigma w_{n}R_{c}\right)
^{3/4}}{3\left( \Delta w\right) ^{3/2}}t^{3/2}.
\end{equation}

If the transition from neutron to quark phase is driven by viscous processes
only, then
\begin{equation}
t_{conv}\approx R_{c}\frac{\left( \frac{\mu _{u}^{4}+\mu _{d}^{4}}{\pi ^{2}}%
-3B-n_{B}m_{n}-\frac{\gamma }{\gamma -1}Kn_{B}^{\gamma }\right) ^{2}}{%
2\sigma \left[ \lambda _{n}T+2\left( \frac{4}{3}\eta _{n}+\xi _{n}\right) %
\right] }R^{2},
\end{equation}
and
\begin{equation}
V(t)\approx \frac{2^{5}\pi \sigma ^{3/2}}{3\left( \Delta w\right)
^{3}R_{c}^{3/2}}\left[ \lambda _{n}T+2\left( \frac{4}{3}\eta _{n}+\xi
_{n}\right) \right] ^{3/2}t^{3/2}.
\end{equation}

\section{Endothermic nucleation of quark matter bubbles in the core of
rotating neutron stars}

In the previous Sections we have considered the process of nucleation of
quark bubbles in neutron star cores, and we have derived the conditions for
quark matter formation in an astrophysical context. In the present Section
we shall apply the results previously obtained to analyze the possibility of
the transition from neutron to quark phase during the spinning down of a
rapidly rotating neutron star. In this case the rotational energy of the
star can be used to trigger the phase transition.

The minimum energy density required to create a quark bubble is
\begin{equation}
u_{q}=\frac{3W_{c}}{4\pi R_{c}^{3}}=\frac{C}{\left( 1+\sqrt{1+b}\right) ^{3}}%
F(b)\left[ 1+\frac{24\pi }{5}\left( \rho _{q}-\rho _{n}\right) ^{2}\frac{%
\sigma ^{2}}{C^{3}}G(b)\right] .  \label{1}
\end{equation}

An estimation of Eq.~(\ref{1}) for $C\approx 20$ MeVfm$^{-3}$, $\gamma
\approx 18$ MeVfm$^{-1}$, $\sigma \approx \left( 75\textrm{ MeV}\right) ^{3}$,
$\rho _{q}\approx -0.4e$ fm$^{-3}$ and $\rho _{n}\approx 0$ gives
$u_{q}\approx 3.76\times 10^{34} ~erg~cm^{-3}$.

We consider that the evolution of the rotating neutron star can be
approximated by a sequence of MacLaurin spheroids \citep{Ch86,ChChZhCh92}. The mass of the star will be denoted by $M$, and the major and
minor axis of the star by $a$ and $c$ (equatorial and polar radius). Then
the eccentricity of the star is defined according to $e=\sqrt{1-\left( \frac{%
c}{a}\right) ^{2}}$. Let $\rho =$constant denote the density of the star,
corresponding to a given value of the angular velocity $\Omega $ and of the
equatorial radius $a$.

Hence the basic equations describing the mass, hydrostatic equilibrium and
gravitational energy $E_{gr}$ of the rotating star are \citep{Ch86}
\begin{equation}
M=\frac{4\pi }{3}a^{3}\rho \left( 1-e^{2}\right) ^{1/2},\Omega ^{2}=2\pi
G\rho f(e),E_{gr}=-\frac{3}{5}\frac{GM^{2}}{a}\frac{\arcsin e}{e},
\end{equation}
where
\begin{equation}
f(e)=\frac{\left( 1-e^{2}\right) ^{1/2}}{e^{3}}\left( 3-2e^{2}\right)
\arcsin e-\frac{3\left( 1-e^{2}\right) }{e^{2}}.
\end{equation}

Assuming that the total mass of the star is a constant, one obtain the
following relation between the variation $\Delta a,\Delta \Omega ,\Delta
\rho $ and $\Delta e$ of the equatorial radius, angular velocity,density and
eccentricity, respectively:
\begin{equation}
\frac{3}{a}\frac{\Delta a}{\Delta \Omega }+\frac{1}{\rho }\frac{\Delta \rho
}{\Delta \Omega }=\frac{e}{1-e^{2}}\frac{\Delta e}{\Delta \Omega }.
\end{equation}

The variation of the eccentricity with respect to the angular velocity is
given by
\begin{equation}
\frac{\Delta e}{\Delta \Omega }=\frac{\Omega -\pi Gf(e)\frac{\Delta \rho }{%
\Delta \Omega }}{\pi G\rho g(e)},
\end{equation}
where
\begin{equation}
g\left( e\right) =\frac{\Delta f(e)}{\Delta e}\approx \frac{df(e)}{de}%
=\allowbreak \frac{2\left( 9-2e^{2}\right) \sqrt{1-e^{2}}+\left(
8e^{2}-9\right) \arcsin e}{e^{4}\sqrt{1-e^{2}}}.
\end{equation}

With the use of the above equations the variation of the gravitational
energy of the neutron star, corresponding to a simultaneous change in
equatorial radius, angular velocity and density, is given by
\begin{equation}
\Delta E_{gr}\left( a,\rho ,\Omega \right) =\frac{3}{5}\frac{GM^{2}}{a}\left[
A\left( e\right) \frac{\Omega }{\pi G\rho }-B\left( e\right) \frac{1}{\rho }%
\frac{\Delta \rho }{\Delta \Omega }\right] \Delta \Omega ,
\end{equation}
where we denoted
\begin{equation}
h(e)=\frac{1}{3}\frac{e}{1-e^{2}}-\frac{1}{e\sqrt{1-e^{2}}}+\frac{\arcsin e}{%
e^{2}},
\end{equation}
\begin{equation}
A(e)=\frac{h(e)}{g(e)},B(e)=\frac{h(e)f(e)}{g(e)}+\frac{1}{3}.
\end{equation}

The gravitational energy density released as the result of the slowing down
of the neutron star can trigger the endothermic phase transition at the
center of the star. Inside a sphere of radius $R_{q}$ the gravitational
energy density is
\begin{equation}
u_{gr}\left( a,\rho ,\Omega \right) =\frac{9}{20\pi }\frac{GM^{2}}{a}%
R_{q}^{-3}\left[ A\left( e\right) \frac{\Omega }{\pi G\rho }-B\left(
e\right) \frac{1}{\rho }\frac{\Delta \rho }{\Delta \Omega }\right] \Delta
\Omega .
\end{equation}

This energy should be greater or equal to the minimum energy necessary for
the formation of a quark bubble, given by Eq.~(\ref{1}). Therefore the
gravitational energy converts to the quark phase the neutron matter inside a
sphere of radius
\begin{eqnarray}
R_{q}\left( a,\rho ,\Omega \right) &=&2.88\left( \frac{M}{M_{\odot }}\right)
^{2/3}\left( \frac{a}{10^{6}}\right) ^{-1/3}\left[ A\left( e\right) \frac{%
\Omega }{\pi G\rho }-B\left( e\right) \frac{1}{\rho }\frac{\Delta \rho }{%
\Delta \Omega }\right] ^{1/3}\times  \nonumber \\
&&\frac{\left( 1+\sqrt{1+b}\right) C^{-1/3}}{\left[ 2+2\left( 1+b\right)
^{3/2}+3b\right] ^{1/3}}\left( \Delta \Omega \right) ^{1/3}\times 10^{6}%
\textrm{ cm.}
\end{eqnarray}

In order the whole neutron star be converted to a quark star the condition $%
R_{q}\left( \rho ,\Omega \right) =a\left( \rho ,\Omega \right) $ must hold.
On the other hand, for typical neutron star densities of the order of $%
10^{14}-10^{15}$ g/$\textrm{cm}^{3}$, the term $A\left( e\right) \Omega /\pi
G\rho $ is much smaller than $B\left( e\right) \left( \Delta \rho /\Delta
\Omega \right) /\rho $,
\begin{equation}
A\left( e\right) \frac{\Omega }{\pi G\rho }<<B\left( e\right) \frac{1}{\rho }%
\frac{\Delta \rho }{\Delta \Omega }.
\end{equation}

For small eccentricities $e$ the function $B(e)$ can be
approximated by $1/3$.

Therefore we obtain the following condition for the relative
change in the central density, necessary to convert a neutron star
to a strange star, due to the endothermic gravitational energy
transfer in the whole volume of the star:
\begin{equation}
\frac{\Delta \rho }{\rho }\approx 0.125\times \frac{C\left[ 2+2\left(
1+b\right) ^{3/2}+3b\right] }{\left( 1+\sqrt{1+b}\right) ^{3}}\left( \frac{M%
}{M_{\odot }}\right) ^{-2/3}.
\end{equation}

For a neutron star of around two solar masses, $M\approx 2M_{\odot }$ and by
assuming $C\approx 20$ MeV and $b\approx 6$, we obtain $\frac{\Delta \rho }{%
\rho }\approx 1.85$.

Of course this result is strongly dependent on the precise numerical values
of $C$ and $b$, which are generally poorly known. $C$ is also a neutron
matter equation of state dependent parameter. For example, assuming for $C$
a value of $10$ MeV will lead to $\frac{\Delta \rho }{\rho }\approx 0.92$.
An increase in the mass $M$ of the star will also decrease the value of the
relative change in the density of the star necessary to convert the neutron
star.

At present, we have very little observational knowledge about how
fast newborn neutron star can rotate. The most well-known young
pulsar is that the Crab pulsar was born with a rotational period
of about $20$ ms \citep{MaTa77}. If so, during its lifetime the
pulsar its central density only increases about $\Delta \rho
_{c}/\rho _{c}\approx 0.001$ \citep{MaXi96}. Currently the fastest
rotating pulsar known is PSR1937+214, which has a period of 1.55ms
or $\Omega \approx 4,000 s^{-1}$. But it has weak magnetic field
and is suggested to be spun-up by accretion \citep{Al82}. However,
some rapidly spinning millisecond pulsars are suggested to be born
by accretion induced collapse from white dwarfs \citep{Ar83}.
Therefore it can not ruled out that some pulsars can be born with
milliseconds and strong magnetic field.

The theoretical investigation of rotating general relativistic
objects performed by \citet{CoShTe94} shows that for some
realistic equations of state of neutron matter this variation of
the central density can be achieved during the complete spin-down
of the star (see Table I).

\begin{table}[h]
\begin{center}
\begin{tabular}{|c|c|c|c|c|}\hline
EOS & $M_{stat}$(in $M_{\odot }$ units) & $R_{stat}$ ($km$) &
$\Delta \Omega $ ($s^{-1}$) & $\frac{\Delta \rho _{c}}{\rho _{c}}$
\\ \hline A & 1.6551 & 8.368 & 10011 & 1.26 \\ \hline AU & 2.1335
& 9.411 & 10587 & 0.97 \\ \hline FPS & 1.7995 & 9.281 & 8874.9 &
0.91 \\ \hline L & 2.7002 & 13.7 & 6482.9 & 1.244 \\ \hline
M&1.8045 & 11.6 & 4437 & 4.189\\ \hline
\end{tabular}
\end{center}
\caption{ Variation of the central density of the rapidly rotating
neutron star for different equations of state
\protect\citep{CoShTe94}: A \citep{Pa71}, AU
\citep{WiFiFa88}, FPS \citep{LoRaPe93}, L \citep{PaPiSm76} and M \citep{PaSm75}. }
\end{table}

The central density increase due to the spin-down can be easily
realized in a short time interval for a special class of stellar type
objects, called magnetars. Magnetars are compact objects with super strong
magnetic fields of the order $B\approx 10^{15}$ Gauss or even higher \citep{Du92,Pa92,ThDu95,ThDu96}. It is now believed that
soft gamma repeaters (SGR's) - a small class (4 confirmed and one candidate)
of high energy transient discovered through their emission of bright X-ray/$%
\gamma $-ray bursts, which repeat on timescales of seconds to
years - are magnetars. There is evidence that the giant SGR flares
involve the cooling of a confined $e^{\pm }$-photon plasma in an
ultrastrong magnetic field \citep{Ly02}. Some authors \citep{Ch98a,Us01} have suggested that SGR's are strange stars, motivated
in part by the super-Eddington luminosities of their giant flares.
The magnetars differ from the canonical pulsars (with low magnetic
fields of the order $B\approx 10^{11}-10^{13}$ Gauss) in the sense
that they spin down much more rapidly.

If we assume that the spin-down of the magnetar is completely
determined by the torque of its relativistic wind emission,
generated via the magnetic dipole radiation, then the time
variation of the angular velocity $\Omega $ of the star is given
by \citet{ShTe83}
\begin{equation}\label{dec}
I\dot{\Omega}=-\frac{2}{3}\frac{\mu ^{2}\Omega ^{3}}{c^{3}},
\end{equation}
where $I$ is the moment of inertia of the star and $\mu =R^{3}B$ is the
magnetic dipole moment. When the star is born with a spin period much
shorter that the observed one, the age of the star is the spin-down age $%
\tau _{sd}=P/\dot{P}$, where $P$ and $\dot{P}$ are the present spin period
and its time derivative. If the spin-down is entirely due to the magnetic
dipole radiation we obtain
\begin{equation}
\tau _{sd}=\frac{\Omega }{\dot{\Omega}}=\frac{3}{2}\frac{Ic^{3}}{%
B^{2}R^{6}\Omega ^{2}}.
\end{equation}

By adopting for the moment of inertia and the radius of the
magnetar the typical values $I=10^{45}$ gcm$^{2}$ and $R=10^{6}$
cm, and by assuming that the star was born with a magnetic field
of the order of  $B\approx 10^{15}$ Gauss and with an angular
velocity of $\Omega _{0}=6000$ rads$^{-1}$, the spin down age is
given by $\tau _{sd}\approx 1125$ s.

From Eq.~(\ref{dec}) it follows that the decay law of the angular velocity $%
\Omega $ of the star is of the form
\begin{equation}
\Omega =\frac{\Omega _{0}}{\sqrt{\frac{4}{3}\frac{B^{2}\Omega
_0^2R^{6}}{c^{3}I}t+1}},
\end{equation}
where $\Omega _{0}$ is the initial angular velocity of the star.

Hence, in a time interval of around
$t\approx 10^{6}$ s and for a magnetic field $B=10^{15}$ G the angular velocity of the star decreases from $\Omega _{0}=6000$ rads$%
^{-1}$ to $\Omega =142$ rads$^{-1}$. For an initial angular
velocity of the order of $\Omega =8000$ rads$^{-1}$ and for a
magnetic field of the order of $B\approx 10^{16} $ G, $\tau
_{sd}\approx 5$ s. In a time interval of around $t\approx 600$ s
the angular velocity of the star decreases to $\Omega \approx 579$
rads$^{-1}$.

Therefore in a magnetar density changes, due to the spin-down,
take place in a short interval of time. Hence they can provide
enough energy to trigger spontaneous nucleation to the quark
phase, once the central density of the star goes above the
critical density.

The nucleation of quark droplets in the presence of strong
magnetic fields essentially depends on the strength of the
magnetic field. The energy of a quantum particle changes
significantly if the magnetic field is of the order or exceeds the
critical value $B_{c}=m^{2}c^{3}/e\hbar $, where $m$ is the
mass of the particle. For electrons the value of the critical field is $%
B_{c}^{(e)}=4.4\times 10^{13}$ G. In order to calculate the value
of the critical magnetic field for quarks we have to use the
estimations for the quark masses. However, there are many
uncertainties in the numerical values of the quark masses, which
play an essential role in the calculation of the critical magnetic
field and which are generally poorly known. In the case of free
quarks, the current quark mass is considered to be in the range of
$m_{q}\sim 5-10$ MeV. Then the corresponding critical magnetic
field is $B_{c}^{(q)}\sim (1-2)\times 10^{2}\times
B_{c}^{(e)}\approx (4.4-8.8)\times 10^{15}$ G. For the current
quark mass, this is the typical strength of the magnetic field at
which the cyclotron lines begin to occur. In this limit the
cyclotron quantum is of the order or greater than its rest energy.
This is equivalent to the requirement that the de Broglie
wavelength is of the order or greater than the Larmor radius of
the particle in the magnetic field \citep{Ch95}. However, from the
physical point of view a better description of the quark mass can
be obtained by using instead the current quark mass the effective
or constitute quark mass, which takes into account the effects of
the strong interactions, and which is of the order $m_q\sim
100-300$ MeV \citep{ChDaLu98}. The critical magnetic field
corresponding to the effective quark mass could be as high as
$B_{c}^{(q)}\sim 10^{17}-10^{18}$ G.

The study of the behavior of the surface and curvature energy
terms in strong magnetic fields have shown that both quantities diverge for $%
B>B_{c}^{(q)}$, with the curvature term diverging much faster.
Consequently, in the presence of strong
magnetic fields the rate of stable quark droplet formation per unit volume $%
I\sim T^{4}\exp \left( -W/T\right) $ tends to zero, $I\rightarrow
0$, and therefore there cannot be any thermal nucleation of quark
droplets in neutron star cores \citep{Ch95,Ch96a}.

The value of $B$ inside a magnetar is not known, and it is not
certain if the interior magnetic field is much stronger than the
surface field. If in the interior of the magnetar the value of the
magnetic field can exceed the critical value $B_{c}^{(q)}\sim
5\times 10^{15}$ G, which represents the most conservative
estimation of the critical magnetic field, corresponding to the
free quark mass $m=5$ MeV, then the rate of droplet formation can
be considerably reduced and the transition from neutron matter to
quark matter could not take place.

On the other hand, by taking into account that the physical mass
of the quark is the effective mass, the corresponding critical
magnetic field could be of the  order $B_{c}^{(q)}\sim
10^{17}-10^{18}$ G, value which could be higher than the magnitude
of the magnetic field inside the magnetars. For magnetic fields
smaller than $B_{c}^{(q)}$, the endothermic nucleation process can
take place in a short time interval during the spin-down of the
magnetar.

\section{Discussions and final remarks}

In the present paper we have considered a kinetic theory of strange matter
nucleation in neutron stars. We derived general necessary conditions for
neutron to quark matter conversion. There have been several assumptions made
in the derivation of the results, mostly related to the form of the growth
speed of the bubble. The result is valid only if non-linear effects can be
ignored, and the linearized hydrodynamic equations are applicable.
Furthermore, we have assumed that the radii of the bubbles are larger than
the correlation length and that heating due to dissipation is slow, causing
the temperature to vary slowly across the bubble wall. We also considered
that the phase transition is strongly first order, releasing considerably
heat.

The growth of the critical size quark bubbles nucleated in the first order
transition in the neutron star core is governed by the dynamical prefactor $%
k $. There are several expressions for the prefactor obtained in the
physical literature. In the Csernai-Kapusta approach \citep{CsKa92}, the
prefactor essentially depends on the dissipative properties (shear, bulk
viscosity and thermal conductivity) of neutron matter. Therefore, in a
non-viscous medium with $\xi _{n}=\eta _{n}=\lambda _{n}=0$, the growth of
the quark bubbles cannot take place. The main physical assumption is that
the energy flow is provided by the viscous effects only.

On the contrary, one can argue that the energy flow does not vanish in
absence of heat conduction or viscous damping \citep{RuFr96,ShMoGu01}%
. In this case, and in the limit of zero viscosity, the prefactor depends
only on two scale parameters, the correlation length $\zeta $ and the
critical radius of the quark bubble $R_{c}$. For a viscous medium the
prefactor can simply be written as the sum of the viscous and non-viscous
term, with the viscous term not affecting the growth process significantly.
The previous (viscous) results can be reobtained by using some assumptions
for the velocity of sound in the medium around the saddle configuration.

The condition for the formation of at least quark bubble in the neutron
matter as a result of the thermal fluctuation is given by Eq.~(\ref{cond1}).
The condition includes both the cases of viscous and ideal neutron matter.
It is extremely sensitive with respect to the numerical values of the
microscopic parameters characterizing the quark bubble, like the surface
tension $\sigma $, the bag constant $B$ or the curvature coefficient $\gamma
$. In discussing the astrophysical implications of the expression (\ref
{cond1}) we shall consider separately the two limiting cases, corresponding
to viscous and non-viscous neutron matter, respectively.

In the case of ideal neutron matter, the condition of the formation of a
quark bubble is given by Eq.~(\ref{cond2}). This equation determines at
which temperature the phase transition takes place. For example, in the case
of the Walecka mean field equation of state \citep{Wa75}, with $K=15758$ and $%
\gamma =4.95$ \citep{LuBeVu94}, the temperature $T_{c}$ necessary for the
formation of a stable quark bubble in a second ($\Delta t=1$ s), in a volume
$\Delta V_{0}$ with radius $\Delta R_{0}=1$ km ($\Delta V_{0}=4\pi /3$ km$%
^{3} $), and at the center of the neutron star, is around $T_{c}=12-13$ MeV,
for $\sigma =(85$ MeV$)^{3}$. Here we have also assumed the standard values
for the quark matter chemical potential and for the bag constant, $\mu
_{u}=\mu _{d}=280$ MeV and $B=60$ MeVfm$^{-3}$ \citep{ChDaLu98}. Since the
temperature of a newborn neutron star is around $10$ MeV \citep{OlMa94}, this
value of the surface tension seems to rule out, for the given equation of
state, the possibility of formation of quark nuclei in the neutron matter.
However, a small variation in the value of $\sigma $, $\sigma =(75$ MeV$%
)^{3} $ reduces the temperature for quark nuclei formation to $T_{c}=4-6$
MeV, a value which does not exclude the possibility that quark nuclei
formation can be initiated in the early stages of the neutron star
evolution. Smaller values of $\sigma $, of the order of $\sigma =(50$ MeV$%
)^{3}$, can lower even more the phase transition temperature, making it
possible even for cold neutron stars. The transition temperature is
relatively insensitive to the details of the equation of state of the dense
neutron matter.

If the temperature at which the phase transition is initiated is strongly
dependent by the microscopic model adopted for describing the neutron matter
properties, the total transition time of the neutron star to a quark star,
given by Eq.~(\ref{conv}), is much less dependent on the details of the
equation of state or the exact numerical values of the surface tension or
curvature of the bubble. Generally, for a neutron star with radius $R=12-15$
km, the conversion time is very long, of the order of $10^{6}$ years for $%
\sigma \approx (75$ MeV$)^{3}$. For $\sigma \approx (50$ MeV$)^{3}$ or even
lower, the conversion time could have values of the order of $10^{7}$ years,
which are still of the same order of magnitude as the lifetime of a neutron
star ($10^{7}$ years). Hence, in the present approach, which is limited to
the consideration of linear effects only in the hydrodynamic description of
the first order phase transitions for an ideal neutron fluid, the growth of
the quark nuclei is very slow. Therefore the results of the present analysis
suggest that even quark nuclei could appear inside the neutron star at a
very early stage, the transition to a pure quark phase ends only in the
final stage of its evolution.

If the neutron matter-quark matter phase transition is mainly driven by
viscous processes, like in the Csernai-Kapusta scenario, then the time
necessary for a neutron star to convert to a quark star follows from
Eqs.~(\ref{k1}) and (\ref{tconv}). In order to numerically estimate the
conversion
time, we need an estimation of viscosity and conduction coefficients of
nuclear matter. We shall also consider two limiting cases for the phase
transition. Below the critical density $\rho =\rho _{c}\approx 10^{14}$ g/cm$%
^{3}$, cold neutron matter consists of two distinct phases. At low
density, neutron star matter consists of nuclei which form a solid
lattice for temperatures below the melting temperature and a sea
of relativistic electrons. When the density reaches $\rho \approx
10^{11}$ g/cm$^{3}$, neutrons begin to ''drip'' from the
neutron-rich nuclei and then, in addition to the relativistic
electrons there is a sea of non-relativistic neutrons. At the
critical density the neutron-rich nuclei dissolve, leaving seas of
degenerate neutrons, protons and electrons. The neutrons in cold
neutron star matter are quite likely superfluid, and therefore the
transport coefficients are dominated by electrons, with the motion
of electron determined by electron-electron and electron-proton
scattering \citep{FlIt79}
As a result the viscosity is much lower, but the thermal conductivity is
only slightly smaller as compared to the case in which both neutrons and
protons form normal fluids \citep{FlIt79}. Hence, we can neglect the
viscosity coefficients with respect to the thermal conductivity and
approximate $\lambda _{n}$ by $\lambda _{n}\approx \lambda _{0}\rho
_{14}/T_{8}$, with $\lambda _{0}\approx 10^{23}$ \citep{FlIt79}. A slightly
different expression for the leading term in thermal conductivity has been
proposed in \citet{Da84}, $\lambda _{n}T\approx \lambda _{0}\left(
n/n_{0}\right) ^{1/4}$, with $\lambda _{0}=0.15$ fm$^{-3}$ and $n_{0}=0.145$
fm$^{-3}$. Hence, by using this last expression, we obtain for $t_{conv}$
\begin{equation}
t_{conv}\approx R_{c}\frac{\left( \Delta w\right) ^{2}}{4\sigma \lambda
_{0}\left( n/n_{0}\right) ^{1/4}}R^{2}.  \label{therm}
\end{equation}

The transition time is independent of the temperature of the neutron matter.
It depends only on the microscopic properties of the quark bubble (surface
tension and curvature coefficient), density and the equation of state of the
matter inside the star. For $\mu _{u}=\mu _{d}=280$ MeV and $B=60$ MeVfm$%
^{-3}$ and for $\sigma \approx (75$ MeV$)^{3}$ Eq.~(\ref{therm}) gives, for
a neutron star with radius $R=12$ km and for several equations of state
values of the conversion time of the order $t_{conv}\approx 10^{11}$ years.

In the case of young and hot neutron stars, the main dissipative mechanism
which could drive the phase transition is the shear viscosity of the neutron
matter. The temperature dependence of the shear viscosity coefficient can be
approximated by $\eta _{n}\approx \left( \eta _{0}/T^{2}\right) \left(
n/n_{0}\right) ^{2}$, with $\eta _{0}=1700$ MeV$^{3}$/fm$^{2}$ \citep{Da84}.
Consequently, by neglecting the heat conduction and the bulk viscosity, the
conversion time becomes
\begin{equation}
t_{conv}\approx \frac{3}{32}R_{c}\frac{\left( \Delta w\right) ^{2}}{\sigma
\eta _{n}}R^{2}\approx \frac{3}{32}R_{c}\frac{\left( \Delta w\right) ^{2}}{%
\sigma \eta _{0}\left( n/n_{0}\right) ^{2}}T^{2}R^{2}.  \label{shear}
\end{equation}

In small temperatures, the shear viscosity
becomes very large. Hence, for a shear viscosity driven phase transition the
conversion time could be very small.

In Fig.~1 we have represented $t_{conv}$ given by Eq.~(\ref{shear})
as a function of temperature, for different equations of state of
neutron matter. We have considered four equations of state,
namely, the equation of state (EOS) of the free neutron gas
\citep{ShTe83}, the Bethe-Johnson EOS \citep{ShTe83}, the
Lattimer-Ravenhall EOS \citep{LaRa78} and the Walecka EOS \citep{Wa75}. In all cases we calculated the conversion time for a
neutron star with radius $R=10$ km.

\begin{figure}
\plotone{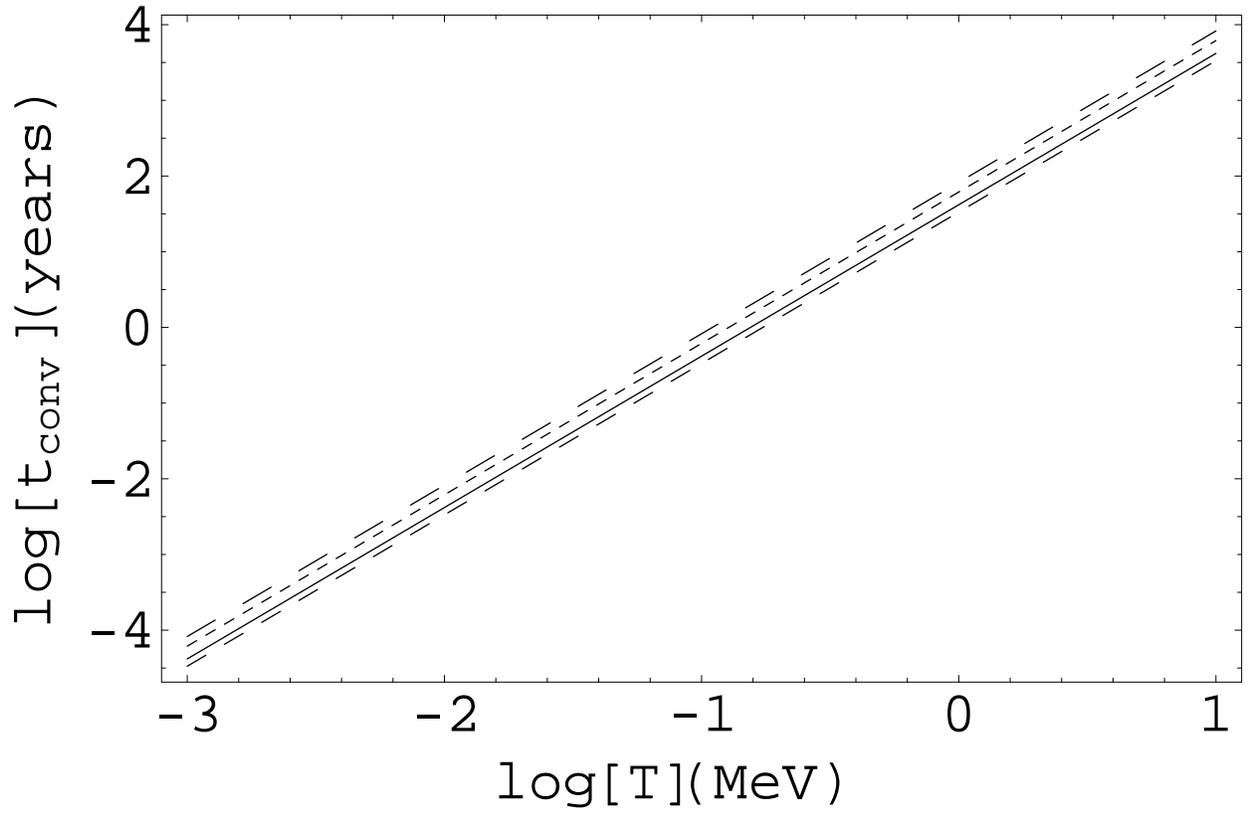}
\caption{Variation of the conversion time $t_{conv}$ for a neutron
star with radius $R=10$ km, for different equations of state of
the neutron matter: free neutrons EOS (solid curve), Bethe-Johnson
EOS (dotted curve), Lattimer-Ravenhall EOS (short-dashed curve)
and Walecka EOS (long-dashed curve). \label{FIG1}}
\end{figure}

As one can see from the figure, in this case the conversion time
is much shorter as compared to the ideal neutron matter case or
thermal conduction driven phase transition. $t_{conv}$ is rapidly
decreasing with the temperature. For a neutron star with rapidly
cooling processes, e.g. direct URCA process, kaon condensation
etc., the stellar temperature can decrease to less than
$10^{-2}$MeV in a time scale of $10^{-4}$ yr \citep{Ts98}.
According to Fig. 1, the conversion can take place in a time scale
less than $10^{-3}$ yr if the spin-down time scale is less than
this time scale. This can be case if this neutron star is a
magnetar. We should point out that the interior of magnetars are
not quantum liquid, i.e. superfluid or superconducting, because
the strong magnetic field can destroy the nucleon cooper pairs
\citep{Ch92}.

On the other hand, if the neutrons and protons are superfluid,
then the transport properties are determined completely by
electron-electron scattering. In fact in this case the viscosity
of the nuclear fluid is given by $\eta ^{-1}=\left( 15\pi
^{4}\alpha ^{2}/2p_{e}^{5}\right) \left( 2p_{e}/k_{ft}\right)
\left[ 5/2+3\left( m_{e}/p_{e}\right) ^{2}+\left(
m_{e}/p_{e}\right) ^{4}\right] T^{2}$ \citep{FlIt76}, where
$\alpha $ is the fine structure constant, $k_{ft}=\left( 4\alpha
p_{e}\epsilon _{e}/\pi \right) ^{1/2}$ is the Thomas-Fermi wave
vector, and $m_e$, $p_{e}$ and $\epsilon _{e}$ are the electron
mass, momentum and energy, respectively.

We assume that the electron gas inside the neutron star is extreme
relativistic and strongly degenerate. Then it follows that
$p_{e}\approx p_{F}=\left( 3n_{e}/8\pi \right) ^{1/3}$
\citep{ShTe83}, with $p_F$ the Fermi momentum and $n_{e}$ the
electron number density. Since $n_{e}$ can be related to the total
particle number $n$ by means of the relation $n_{e}=Y_{e}n$, with
$Y_{e}$ the mean number of electrons per baryon, it follows that
$p_{e}\approx \left( 3Y_{e}n_{0}/8\pi \right) ^{1/3}\left(
n/n_{0}\right) ^{1/3}$. Hence we obtain for the shear viscosity
coefficient of the nuclear matter
\begin{equation}
\eta^{-1}\approx \frac{15}{2}\frac{8^{5/3}\pi^{37/6}\alpha ^{3/2}}{\left(
3Y_{e}n_{0}\right) ^{5/3}\left( n/n_{0}\right) ^{5/3}}\left\{ \frac{5}{2}+3%
\left[ \frac{m_{e}\left( 8\pi \right) ^{1/3}}{\left( 3Y_{e}n_{0}\right)
^{1/3}\left( n/n_{0}\right) ^{1/3}}\right] ^{2}+\left[ \frac{m_{e}\left(
8\pi \right) ^{1/3}}{\left( 3Y_{e}n_{0}\right) ^{1/3}\left( n/n_{0}\right)
^{1/3}}\right] ^{4}\right\} T^{2}.
\end{equation}

The variation of the conversion time, as a function of temperature,
is represented, for a superfluid neutron and proton core of the star with $Y_e=0.1$, in Fig.~2.

\begin{figure}
\plotone{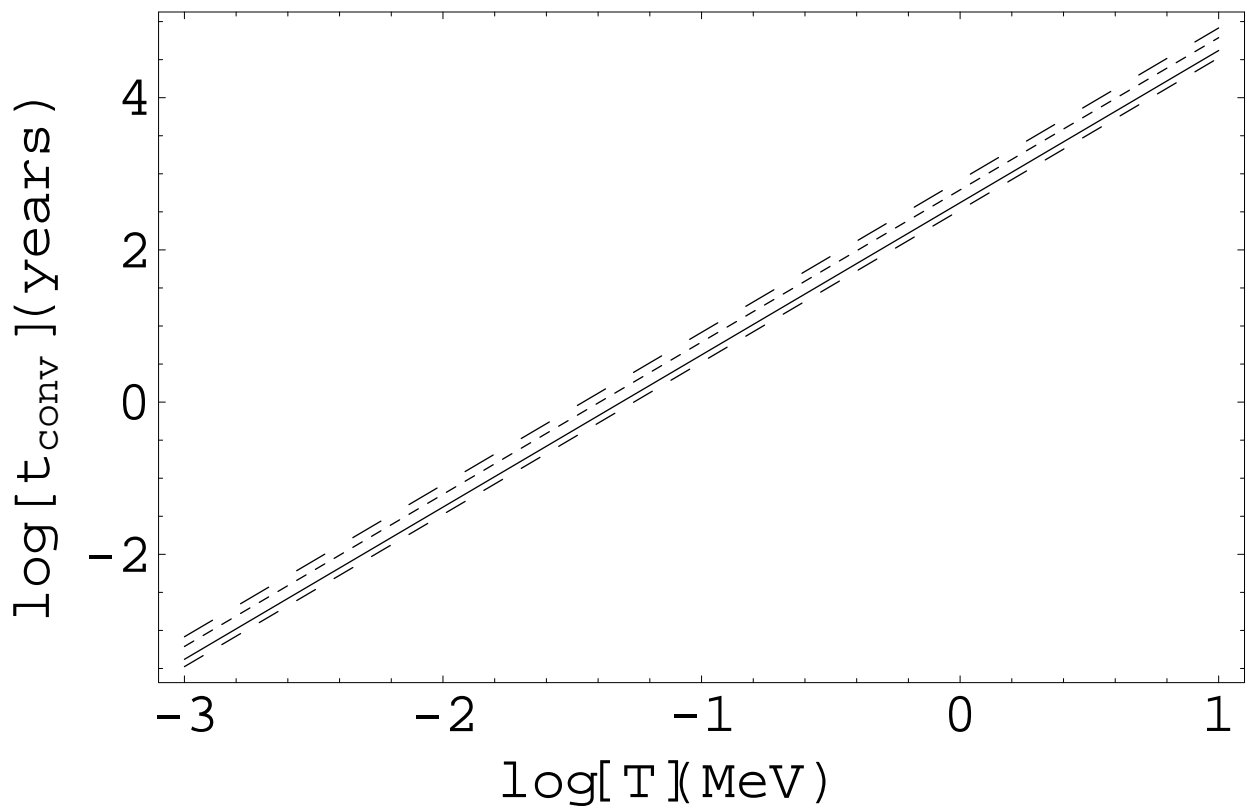}
\caption{Variation of the conversion time $t_{conv}$ for a neutron
star with radius $R=10$ km to a quark star, in the case of the
presence of a superfluid core, for different equations of state of
the neutron matter: free neutrons EOS (solid curve), Bethe-Johnson
EOS (dotted curve), Lattimer-Ravenhall EOS (short-dashed curve)
and Walecka EOS (long-dashed curve). \label{FIG2}}
\end{figure}

In this case, since the viscosity of the superfluid nuclear matter
is much smaller, the conversion time
is longer than for a nuclear fluid dominated by the neutron viscosity. However, in this
case the phase transition from neutron matter to quark matter can also take place at very
low temperatures. Consequently, a very short $t_{conv}$, of the order
of hours or a few days, is also allowed in the framework of this model.

It is widely accepted that at asymptotic densities the ground
state of QCD with $m_s=0$ is the color-flavor locked (CFL) phase
\citep{Al98,Ra01}. In this phase, color gauge symmetry is
completely broken, as are both chiral symmetry and baryon number.
The effective coupling is weak and the low-energy properties can
be determined by using methods from the theory of
superconductivity. For large gaps $(\Delta \sim 100MeV)$, the CFL
phase is rigorously electrically neutral, despite the unequal
quark masses, and even in the presence of electron chemical
potential. The transition from nuclear matter to quark matter via
bubble nucleation may be greatly simplified if the transition
occurs directly to quark matter in the CFL phase \citep{Ra01}. For
a given baryonic chemical potential $\bar{\mu}$, electrically
neutral nuclear matter and electrically neutral quark matter have
different values of the electronic chemical potential $\mu _{e}$.
Since $\mu _{e}$ must be continuous across any interface, a mixed
phase region is formed, within which positively charged nuclear
matter and and negatively charged quark matter with the same $\mu
_{e}$ coexist at any given radius. The growth of the quark
droplets is at the expense of the nuclear matter. However, if the
quark matter is in the CFL phase, an interface between bulk
nuclear matter with non-zero electron number $N_{e}\neq 0$ and CFL
quark matter with $N_{e}=0$ may be stable, as long as $\mu _{e}$
satisfies the condition $\left| m_{s}^{2}/4\bar{\mu}-\delta \mu
\right| <\Delta /\sqrt{2}$, with $\delta \mu $ the variation of
the chemical potential \citep{Ra01}. Therefore, the existence of
the CFL phase supports the idea that in neutron stars quark matter
and baryonic matter may be continuously connected. Then the growth
of the interface between the neutron and CFL quark phase could be
the result of exterior energy absorption, via an endothermic
mechanism.

In summary, by using the kinetic nucleation theory we have
considered the possibility that the neutron matter-quark matter
phase transition could be triggered by the gravitational energy
released during the spin-down of a pulsar. This process could take
place in a very short time interval, of the order of few tens
seconds, in a class of dense compact objects, with very high
magnetic fields, called magnetars. However, the percentage of
neutron star matter that can be converted into quark matter
depends on the initial period, the neutron star mass and on the
equation of state of the neutron and quark phases. An important
and open issue is how could we determine if such a process is
taking place or took already place in some pulsars or magnetars,
and what are the observational signatures of the endothermic
neutron matter-quark matter transition. In particular, it would be
interesting to know if such a process could be a possible
gamma-ray burst mechanism, as suggested by \citet{Be03}. All of
these issues will be considered in a subsequent paper
\citep{Ta04}.

\acknowledgments This work is supported by a RGC grant of the Hong
Kong government of the SAR of China. The authors would like to
thank to the anonymous referee for comments that helped to improve
the manuscript.

\end{document}